\documentclass[letterpaper,twocolumn,10pt]{article}
\usepackage[hyphens]{url}
\usepackage{usenix-2020-09}

\usepackage{xcolor}
\usepackage{subcaption}
\usepackage{xspace}
\usepackage{listings}
    \lstdefinelanguage{PROMELA}
    {
        morekeywords={if,fi,do,od,proctype,init,byte,bool,atomic},
        otherkeywords={->,::},
        sensitive=false,
        basicstyle=\scriptsize\ttfamily,
        frame=single,
        xleftmargin=\parindent,
        escapeinside={\%*}{*)},
    }
\usepackage{multirow}
\usepackage{latexsym}
\usepackage{amssymb}
\usepackage{pifont}

\newcommand{\spin}{\textsc{Spin}\xspace}
\newcommand{\promela}{\textsc{Promela}\xspace}
\newcommand{\hide}[1]{}
\newcommand{\bred}[1]{\textcolor{red}{#1}} 

\newcommand{\blind}[1]{Anonymous GitHub Repository.}
\newcommand{\cmark}{\ding{51}}
\newcommand{\xmark}{\ding{55}}

\begin{document}

\date{}

\title{\Large \bf Analyzing FreeRTOS Scheduling Behaviors with the Spin Model Checker}


\author{
{\rm Chen-Kai Lin}\\
Academia Sinica, Taiwan
\and
{\rm Bow-Yaw Wang}\\
Academia Sinica, Taiwan
}

%

\maketitle

\begin{abstract}
  FreeRTOS is a real-time operating system with configurable
  scheduling policies.
  Its portability and configurability make FreeRTOS one of the most
  popular real-time operating systems for embedded devices.
  We formally analyze the FreeRTOS scheduler on ARM Cortex-M4
  processor in this work. Specifically, we build a formal model
  for the FreeRTOS ARM Cortex-M4 port and apply model checking to find
  errors in our models for FreeRTOS example applications.
  Intriguingly, several errors are found in our application models
  under different scheduling policies. In order to confirm our
  findings, we modify application programs distributed by FreeRTOS and
  reproduce assertion failures on the STM32F429I-DISC1 board.
\end{abstract}

\section{Introduction}


FreeRTOS is an open-sourced real-time operating system
supporting various architectures~\cite{web:FreeRTOS}. Because of its
portability, FreeRTOS is the third most popular operating system in
2019~\cite{2019EmbeddedMarket}. It has been used in numerous embedded
devices~\cite{7793001, 7943952, Morris2016, Clagett2017DellingrNG,
Leppinen2014, 8861596, surrey26828}.


One of the most useful features in FreeRTOS is multi-tasking on
uni-processor embedded devices. Through multi-tasking, 
applications can be divided into several simpler tasks sharing
processor time. Indeed, FreeRTOS provides three configurable
scheduling policies for different applications. Multi-tasking
nevertheless can induce undesirable phenomenon such as
deadlocks or starvation. It is therefore crucial to 
prevent such errors in deployment. Multi-tasking
errors on the other hand are notoriously evasive. Due to complex
interleavings among tasks, a very limited number of
system behaviors can be tested. In order to check multi-tasking
in FreeRTOS more thoroughly, we apply model
checking in our analysis.


Model checking is a formal technique to analyze properties about
systems~\cite{10.5555/332656}. In model checking, behaviors of the
system under verification are specified in a formal model. Model
checkers can verify the model automatically with formal
properties provided by users. Different from testing tools, model
checkers search model behaviors exhaustively. If a deviant behavior is 
found, it is reported to verifiers. If no deviance can be found after
exhaustive search, all model behaviors conform to the specified
property. The formal model is thus verified.

In this paper, we develop formal models for the FreeRTOS scheduler on
ARM Cortex-M4 processors and analyze its properties by the \spin model
checker. Based on the reference manual, we build formal
models for the ARM Cortex-M4 interrupt handling mechanism.
Particularly, optimizing mechanisms such as tail chaining are
implemented in our models. Through examining source codes of the
FreeRTOS ARM Cortex-M4 port, formal models for the FreeRTOS scheduler,
thread-safe data structures, and its applications are also
built. 
Particularly, all FreeROTS scheduling policies are formalized in our
models as well.

With our behavior models for FreeRTOS, it remains to identify formal
properties to check. Such properties however can be tricky to find.
For formal analysis, high-level informal properties such as absence of
deadlocks or starvation need to be specified concretely. In complex
systems like FreeRTOS, high-level properties are often asserted with
caveats to preclude minor or unrealistic errors. It can be very
tedious to specify caveats formally. Moreover, one can not be sure of
these caveats without FreeRTOS developers' help. Different developers
can also have different views on properties and caveats. Subsequently,
formal properties specified by verifiers can be contrived or even
incorrect.

We solve the property specification problem by verifying example
applications in the FreeRTOS distribution. In order to highlight
FreeRTOS features, developers provide a number of example 
applications for demonstration. Most example applications contain
assertions to specify expected behaviors during execution.
Intuitively, no assertion failure should be observed on any
multi-tasking execution. We therefore add assertions to our model and
verify them with the \spin model checker. Intriguingly, the model
checker reports errors on several example application models. 


Assertion errors found in formal analysis do not necessarily imply
assertion failures in real execution. In order to confirm our
findings, we modify FreeRTOS example applications to reproduce error
traces in our 
formal behavior models. If assertion errors in formal analysis are
genuine, we should observe assertion failures on real hardware. Using
the STM32F429I-DISC1 board from STMicroelectronics, we successfully
reproduce assertion failures in our experiments. We use the remote
GDB debugger to confirm failures at intended assertions.
All assertion failures require delicate interactions among
tasks, the FreeRTOS scheduler, and the ARM Cortex-M4 interrupt
handling mechanism. They are unlikely to be discovered by testing.

This paper is organized as follows. Section~\ref{sec:background}
introduces the \spin model checker. Section~\ref{sec:methodology}
presents analysis methodology. Section~\ref{sec:ExUnits} describes
a task and an interrupt handler models. Section~\ref{sec:scheduler}
describes FreeRTOS scheduling policies. Section~\ref{sec:delay:suspend}
describes tasks delaying and suspending mechanism.
Section~\ref{sec:thread-safe-data} describes a thread-safe data
shared by tasks. Section~\ref{sec:applications} describes FreeRTOS
example applications. Section~\ref{sec:formal-properties} classifies
assertions in the applications as properties. Section~\ref{sec:evaluation}
reports verification results and discussion.
Section~\ref{sec:relate} gives related works.
Section~\ref{sec:conclusion} concludes.

\section{Background}
\label{sec:background}
\hide{
In order to analyze a multitude of execution scenarios with FreeRTOS
schedulers, we apply formal verification in this work. Formal
verification generally refers to system analysis techniques based on
mathematical or logical reasoning. Different from testing, formal
techniques aim to prove the correctness of systems by exploring all
possible system behaviors. If no problem is found after such
exploration, it is logically impossible to have any error under
assumptions imposed by verifiers.
}

Model checking is an automatic formal verification technique. In model
checking, systems under verification are specified as formal
models. Properties about systems are also formalized by logical
properties about formal models. Given a formal model and a logical
property, a model checker automatically verifies the logical property
against the model through mathematical reasoning. If the model is
verified, the property holds in the model mathematically. If the model
is not verified, the model checker returns a trace to witness the
error.

\hide{
An important application of model checking is to verify
control-intensive concurrent systems. In such systems, a number of
concurrent processes interacting one another. \emph{All} interactions
must be analyzed in order to prove the correctness of
systems. Interactions nevertheless can be intricate and tedious for
manual analysis. Testing on the other hand is unlikely to explore all
interactions thoroughly. Indeed, model checking has established its
early fame in analyzing network protocols and digital circuits.
}

\spin is a model checker designed for analyzing communicating
concurrent processes~\cite{10.1109/32.588521}. It offers 
the \promela (PROcess MEta LAnguage) language to specify
formal models for systems. A formal model in \promela consists of the 
\emph{main} process. Additional processes can be instantiated if
needed. A process contains a sequence of \emph{commands}.
Commands must be \emph{enabled} before execution. Enabled commands in
all processes (including the main process) are executed
interleavingly. That is, exactly one enabled command is executed at
any time. If several commands from different processes are enabled, 
one of the enabled commands is executed non-deterministically. 
If there is no enabled command among all processes, it is a
\emph{deadlock}. The \promela language allows verifiers to specify
assertions in processes. An \emph{assertion} command contains a
Boolean expression and is always enabled. Its
Boolean expression is evaluated when an assertion command is
executed. If the Boolean expression is false, it is an assertion error.
Recall that enabled commands are executed non-deterministically.
Non-deterministic executions result in different \emph{traces}.
Some traces have assertion errors but others have not.

Since traces formalize system behaviors, a deadlock or an assertion
error in any trace represent undesirable system behaviors. We
therefore would like to check if deadlocks or assertion errors
occur among all traces. The \spin model
checker systematically explores all traces with sophisticated
algorithms. If a deadlock or an assertion error occurs in any trace,
\spin will find the \emph{error trace} and report it as a witness. If
the model 
checker does not find any deadlock or assertion error after
exploring all traces, the model is verified. 

\hide{
is designed for verifying concurrency problems of
a (multithreading) program. Spin verifies an abstract model against some given
invariants translated from the concurrency problems. The abstract model supported
by SPIN is programmed in PROMELA, an acronym for \textit{Process Meta Language}.
In PROMELA, a process is an abstraction of a thread, which accesses shared resources
between other threads in non-deterministic order.
By exploring interleaving steps between processes exhaustively, Spin verifies
a model against some invariants, otherwise it reports the trail indicating
violation of invariants on the model.
}

\begin{figure}
\begin{lstlisting}[language=PROMELA]
byte counter;           bool b[3];
proctype Guess(byte idx) {
  atomic {
    if
    :: true -> b[idx] = false; counter++;
    :: true -> b[idx] = true; counter++;
    fi;
} }
init {
  counter = 0;
  run Guess(0);    run Guess(1);    run Guess(2);
  do
  :: counter == 3 -> break
  od;
  assert (!( b[0] && b[1] && !b[2] ));
}
\end{lstlisting}
\caption{Boolean Satisfiability Solver}
\label{fig:sat}
\end{figure}

Figure~\ref{fig:sat} gives a simple \promela model solving Boolean
satisfiability from the \spin distribution. It contains the variable
\texttt{counter} and the Boolean 
array \texttt{b} of size 3. The model declares a process
called \texttt{Guess}. Given an index $\mathit{idx}$,
\texttt{Guess($\mathit{idx}$)} assigns a Boolean 
value to the array element with the index $\mathit{idx}$.
The \texttt{\textbf{if}} command contains two
commands. Since both commands are enabled, one of them is chosen to
assign the array element non-deterministically. The variable
\texttt{counter} is then incremented by one. The
\texttt{\textbf{atomic}} keyword indicates all commands in its
brackets are executed atomically.

The keyword \texttt{\textbf{init}} designates the main process. In the
main process, \texttt{counter} is set to zero. An instance of the
\texttt{Guess} process is created for each element in the Boolean
array. After process instantiation, the main process enters a busy-waiting
loop. The \texttt{\textbf{do}} command executes an enabled command
repeatedly until the \texttt{\textbf{break}} command. The main process
subsequently waits until the value of \texttt{counter} is equal to 3. 
When \texttt{counter} is 3, all array elements have been assigned.
The main process checks if the Boolean expression in
the \texttt{\textbf{assert}} command is true. If not, it is an
assertion error.

Recall that the \texttt{Guess} process assigns a Boolean value to an
array element non-deterministically. Depending on non-deterministic
assignments, an assertion error may or may not occur in the
\texttt{\textbf{assert}} command. The \spin model checker 
verifies if an assertion error occurs in all traces induced by
different non-deterministic assignments. In the example, \spin finds an
assertion error and reports the trace ending with \texttt{b[0]} =
\texttt{b[1]} = \texttt{true} and \texttt{b[2]} = \texttt{false}.

Observe that assignments in the \texttt{Guess} process are not the
only non-deterministic behaviors. Since the main process and three
instances of the \texttt{Guess} process are running, the order of
execution is also non-deterministic. For instance, any of the three
\texttt{Guess} processes may assign to its array element before the
other two. In Figure~\ref{fig:sat}, several traces indeed end with the
state where the assertion error occurs. The \spin model checker
reports the first assertion error found.

\hide{
A toy model: proctype, if/fi, do/od, double colons, atomic, the guard statement \texttt{->},
if no option is true, the conditional statement is blocked, an expression (EP==id),
Spin cannot verify dynamic memory allocation,
}

In addition to assertions, \spin allows verifiers to specify
properties with Linear Temporal Logic (LTL)
formulas. Particularly, we will use the LTL formula
$\Box\Diamond\mathit{Loc}$ where $\mathit{Loc}$ denotes a process
location. A trace satisfies $\Box\Diamond\mathit{Loc}$ if
it visits the process location $\mathit{Loc}$ infinitely many
times. We use the formula $\Box\Diamond\mathit{Loc}$ to specify that a
process is free of starvation. More concretely, let $\mathit{Loc}$ be
the location where a process finishes its job. A trace satisfies
$\Box\Diamond\mathit{Loc}$ if the process finishes its job
infinitely many times. 
More generally, 
the LTL formula $\Box\Diamond\mathit{Loc_0} \rightarrow
\Box\Diamond\mathit{Loc_1}$ specifies that the process location
$\mathit{Loc_1}$ is visited infinitely many times if $\mathit{Loc_0}$
is visited infinitely many times.

\section{Methodology Overview}
\label{sec:methodology}

In order to support different architectures, the FreeRTOS scheduler
contains both architecture-dependent and
architecture-independent codes. Roughly, scheduling policies are 
independent of underlying architectures. They provide abstract
programming models for applications. Their implementations
must depend on interrupt mechanisms in underlying architectures
however. 
For instance, the FreeRTOS scheduler is called during periodic and
sporadic interrupts in the ARM Cortex-M4 port. For analysis,
it is essential to consider as many interrupt sequences as
possible. Generating such interrupt sequences for testing is
infeasible. A more effective technique is needed.

In this work, we develop a \promela model for the interrupt mechanism
on ARM Cortex-M4 processors. Behaviors of optimizing mechanisms for ARM 
Cortex-M4 processors are carefully formalized in our model. More
importantly, non-deterministic interrupts allow us to explore a
gigantic amount of interrupt sequences unattainable by 
testing.

On top of our formal model for the ARM Cortex-M4 interrupt mechanism,
we then specify a \promela model for the architecture-independent codes
in the FreeRTOS scheduler. All three FreeRTOS scheduling policies are
specified in our model. Our formal model for the FreeRTOS scheduler on
ARM Cortex-M4 processors enables extensive analysis on task
synchronization -- task delay and suspension. We moreover
build formal models for thread-safe data structures such as queues and
locks (send/receive and give/take). These structures are widely used
by FreeRTOS applications.

With formal models, we proceed to verify properties about the FreeRTOS
scheduler. Although abstract properties such as the absence of
deadlock and starvation are easily said, they are not precise enough
for formal analysis. Additionally, properties are unlikely to be
satisfied without provisions. Without FreeRTOS developers' inputs,
contrived or even misleading properties can be verified
meaninglessly. 

We address this problem by verifying FreeRTOS example
applications. Similar to most open-sourced projects, FreeRTOS provides
example applications to illustrate its features. These applications
contain assertions to specify expected behaviors. Intuitively,
these assertions are but formal properties written by FreeRTOS
developers. No assertion failure should be observed under all
circumstances. In order to verify assertions in FreeRTOS example
applications, we build their formal models and check if an assertion
error might occur. Intriguingly, several
assertion errors were found in our analysis.

It is important to recall that formal models are different from real
hardware and software by definition. Assertion errors found on the
models do not necessarily correspond to assertion failures on real
systems. In order to support our findings, we examine the error traces
found by the \spin model checker and reproduce them on the
STM32F429I-DISC1 board. We moreover use the remote debugger GDB to
confirm assertion failures on the ARM Cortex-M4 board. Many
errors found by our formal analysis are successfully realized on real
systems. These assertion failures require intricate
interrupt events. They are unlikely to be found by traditional testing.

\section{Execution Units}
\label{sec:ExUnits}

Our goal is to develop \promela models for the ARM Cortex-M4 interrupt
mechanism, the FreeRTOS scheduler, thread-safe data structures, and
example applications. An application has a number of  
tasks to be executed by the processor. When an interrupt is triggered,
its interrupt handler will be executed by the processor. We therefore
say a task or an interrupt handler are \emph{execution units}.
In our \promela model, an execution unit is formalized as a \promela
process. Commands in the process thus specify the computation
of the execution unit.

\subsection{Task}

Typical FreeRTOS tasks loop forever and never terminate. Their models
are \promela processes with infinite loops (Figure~\ref{fig:task}).
In \promela, an enabled command is executed non-deterministically
among all such commands in all processes. Our FreeRTOS task models
however need to be scheduled by our formal scheduler model before
execution. To this end, we define the global variable
\texttt{EP} (for Executing Process) and assign each execution unit a
unique identification number. Every guard and command in task models
are annotated with the condition \texttt{EP} \texttt{==}
$\mathit{id}$. The FreeRTOS scheduler model in turn assigns the
variable \texttt{EP} to elect task models.

More precisely, each guard $\mathit{bexp}$ is annotated
with the macros 
\texttt{SELE(}$\mathit{id}$\texttt{,}$\mathit{bexp}$\texttt{)} or
\texttt{ELSE(}$\mathit{id}$\texttt{,}$\mathit{bexp}$\texttt{)};
each command $\mathit{cmd}$ is annotated with
\texttt{AWAIT(}$\mathit{id}$\texttt{,}$\mathit{cmd}$\texttt{)}.
The macros \texttt{SELE(}$\mathit{id}$\texttt{,}$\mathit{bexp}$\texttt{)}
and \texttt{ELSE(}$\mathit{id}$\texttt{,}$\mathit{bexp}$\texttt{)} add the
condition \texttt{EP} \texttt{==} $\mathit{id}$ in conjunction with the Boolean
expression $\mathit{bexp}$ and its negation respectively. If the task
model is not scheduled for execution, the variable \texttt{EP} is not
equal to its identification number and the annotated guard is
false. Otherwise, the variable \texttt{EP} is set to its
identification number and the guard $\mathit{bexp}$ is then
checked. The macro
\texttt{AWAIT(}$\mathit{id}$\texttt{,}$\mathit{cmd}$\texttt{)} adds
the guard \texttt{EP} \texttt{==} $\mathit{id}$
to the command $\mathit{cmd}$. The annotated command is enabled
precisely when the task model is scheduled.

\begin{figure}
\begin{lstlisting}[language=PROMELA,mathescape=true]
#define SELE($\mathit{id}$, $\mathit{bexp}$) ( EP==$\mathit{id}$ && ($\mathit{bexp}$) )
#define ELSE($\mathit{id}$, $\mathit{bexp}$) ( EP==$\mathit{id}$ && !($\mathit{bexp}$) )
#define AWAIT($\mathit{id}$, $\mathit{cmd}$) atomic { EP==$\mathit{id}$ -> $\mathit{cmd}$ }
proctype task() {
do
:: SELE($\mathit{id}$, $\mathit{bexp}_0$) -> AWAIT($\mathit{id}$, $\mathit{cmd}_{00}$); ... AWAIT($\mathit{id}$, $\mathit{cmd}_{0n_0}$);
:: SELE($\mathit{id}$, $\mathit{bexp}_1$) -> AWAIT($\mathit{id}$, $\mathit{cmd}_{10}$); ... AWAIT($\mathit{id}$, $\mathit{cmd}_{1n_1}$);
:: ...
:: SELE($\mathit{id}$, $\mathit{bexp}_k$) -> AWAIT($\mathit{id}$, $\mathit{cmd}_{k0}$); ... AWAIT($\mathit{id}$, $\mathit{cmd}_{kn_k}$);
od }
\end{lstlisting}
\caption{A Task Model}
\label{fig:task}
\end{figure}

\subsection{ARM Cortex-M4 Interrupt Handler}

For ARM Cortex-M4 processors, an interrupt is triggered when it is set
to the \emph{pending} state. When an interrupt is pending, the
processor decides 
whether the current execution should be interrupted. If the pending
interrupt has a priority over the current execution \emph{and} is
unmasked, the current execution is interrupted by the
interrupt handler of the pending interrupt. 

Bookkeeping is needed when the current execution is interrupted.
At \emph{exception entry}, the current processor state is pushed onto
a stack, the pending state of the interrupt is unset, and the
processor is prepared to execute the interrupt handler. The interrupt
handler is executed during the \emph{exception taken} phase. When
the interrupt handler finishes its execution, the interrupted
processor state is restored at \emph{exception return}. The
interrupted execution is then resumed.

Recall that a triggered interrupt remains pending when it does not
have the priority or is masked. At exception
return, the processor checks if there is any pending interrupt with a
priority over the topmost processor state on the stack. If so,
exception return is bypassed and the pending interrupt proceeds to
exception entry. This optimization is called \emph{tail chaining}.

\hide{
A distinct feature of our formalization is to give a detailed model
for ARMv7-M interrupts. Indeed, interrupts are crucial in
implementing FreeRTOS scheduling policies. Interrupt handling
mechanisms moreover vary between architectures. In order to analyze
FreeRTOS correctly, it is important to model interrupts carefully in
our formalization. In this subsection, we briefly describe the
interrupt handling mechanism in the ARMv7-M architecture and
illustrate how it is modeled in \promela.
}

Figure~\ref{fig:isr} gives the outline of an interrupt handler model.
An interrupt handler model with the identification number $id$
consists of an infinite loop with a sequence of commands. The macro
\texttt{IRQ($id$)}
(for interrupt request) checks the interrupt conditions before
exception entry. In the macro, \texttt{IS\_PENDING($id$)} and
\texttt{IS\_MASKED($id$)} check whether the
interrupt $id$ is pending and masked respectively.
\texttt{PRIORITIZING(}$id$\texttt{, EP)} checks if the interrupt $id$ has
the priority over the running execution unit \texttt{EP}.
\texttt{IRQ($id$)} uses another macro
\texttt{ExpEntry(}$id$\texttt{)} if all conditions are satisfied.
In \texttt{ExpEntry(}$id$\texttt{)}, the current execution unit
identification number is pushed on a stack, the pending state of the
interrupt $id$ is unset, and the variable \texttt{EP} is assigned to the
identification number $id$ of the interrupt handler model. Commands
in the interrupt handler model can then be executed.
After the interrupt handler model finishes, the \texttt{ExpReturn()}
macro pops an identification number from the stack and
assigns it to the variable \texttt{EP}. The interrupted execution unit
can then continue.


\begin{figure}
\begin{lstlisting}[language=PROMELA,mathescape=true]
#define ExpEntry($id$) PUSH(EP); UNSET_PENDING($id$); EP=$id$
#define ExpReturn() $\,\,$EP=POP()
#define IRQ($id$)                            $\,\,$\
  if                                     $\,\,$\
  :: IS_PENDING($id$) && !IS_MASKED($id$) &&     $\,\,$\
     PRIORITIZING($id$, EP) -> ExpEntry($id$); \
  fi
proctype interrupt_handler() {
  do
  :: atomic { IRQ($id$) };
     AWAIT($id$, $\mathit{cmd}_0$); AWAIT($id$, $\mathit{cmd}_1$); ...
     AWAIT($id$, ExpReturn());
  od }
\end{lstlisting}
\caption{An Interrupt Handler Model}
\label{fig:isr}
\end{figure}

When an interrupt is pending but masked, \texttt{IS\_MASKED($id$)}
is \texttt{true}. The interrupt handler model will not
execute \texttt{ExpEntry(}$\mathit{id}$\texttt{)}. Instead, the
model waits until it becomes unmasked. Whenever
a pending interrupt is unmasked, the corresponding handler model
can check if it should execute \texttt{ExpEntry(}$\mathit{id}$\texttt{)}.

Finally, an interrupt may remain pending when it does not have the
priority over the running execution unit
(\texttt{PRIORITIZING(}$\mathit{id}$\texttt{, EP)} is
\texttt{false}). When the interrupt with the priority finishes, the
interrupt handler will be taken without exception entry by tailing
chaining. Our interrupt handler model also performs tail chaining when
an interrupt is pending due to insufficient priorities (not shown in
Figure~\ref{fig:isr}).

\hide { 

In order to improve performance, the ARMv7-M architecture further
optimize the interrupt handling mechanism by \emph{tail-chaining}.
Consider an interrupt is pending while an interrupt handler is running
by the processor. When the interrupt handler finishes, the processor
will restore the interrupted processor state from the stack in the
Exception return phase and resume the interrupted execution.
If the pending interrupt has the priority, the resumed execution
will be interrupted immediately. The processor will save the processor
state on the stack and execute the handler of the pending interrupt.
Note that the processor state is unnecessarily restored and saved on the
stack. Tail-chaining is developed to avoid such redundancy.
At the Exception return phase, the processor checks whether there is
any pending interrupt with priorities over the topmost execution on
the stack. If so, exception return is bypassed and the handler of the
pending interrupt enters the exception taken phase immediately by tail
chaining. The topmost processor state remains intact as if it is
restored and then saved on the stack. Redundant exception phases
are removed.

} 


\section{FreeRTOS Scheduler}
\label{sec:scheduler}

The FreeRTOS scheduler implements three scheduling policies. 
In \emph{cooperative scheduling}, a running task has to yield the
processor explicitly. In \emph{preemptive scheduling without
  time slicing}, a running task can be preempted by tasks with higher
priorities. Finally, a task can moreover be preempted by using up its
time slice in \emph{preemptive scheduling with time slicing}.
Depending on the policy, the FreeRTOS scheduler is called to elect the
next task at different occasions.

In the FreeRTOS ARM Cortex-M4 port, scheduling policies are
implemented via two interrupt handlers. The interrupt handler for the
software interrupt \emph{PendSV} is used for task scheduling and
context switching. More precisely, the PendSV interrupt handler calls
the FreeRTOS scheduler to elect the next task for execution. After a
task is elected, the scheduler sets up the processor state through the
ARM Cortex-M4 interrupt handling mechanism for context switching. 
After exception return, the context of the newly elected task is
restored. The elected task resumes its execution as if it were
returned from an interrupt handler in FreeRTOS. The PendSV interrupt
is triggered whenever a task needs to be elected in all scheduling
policies.

For preemptive scheduling with time slicing, the PendSV interrupt
needs to be triggered at every time slice periodically. This is
implemented with the \emph{SysTick} interrupt. The SysTick interrupt
is triggered by a hardware clock periodically. If preemptive
scheduling is enabled, the SysTick interrupt handler triggers the
PendSV interrupt. When the SysTick interrupt handler finishes, the
PendSV interrupt handler will be executed directly by tail chaining.


\begin{figure}
  \begin{lstlisting}[language=PROMELA,mathescape=true]
proctype PendSV_handler() {
  do
  :: atomic { IRQ(PID$_\mathtt{PendSV}$) } ->
       AWAIT(PID$_\mathtt{PendSV}$, SET_TOP(NextTaskId()));
       AWAIT(PID$_\mathtt{PendSV}$, ExpReturn());
  od }
proctype SysTick_handler() {
  do
  :: atomic { IRQ(PID$_\mathtt{SysTick}$) } -> /* Process delayed tasks */
#ifdef PREEMPTIVE_SCHEDULING
       /* SET_PENDING(PID$_\mathtt{PendSV}$) if the expired tasks have
        * the priority over the running task. */
#ifdef TIME_SLICING
       /* SET_PENDING(PID$_\mathtt{PendSV}$) */
#endif
#endif
       AWAIT(PID$_\mathtt{SysTick}$, ExpReturn());
  od }
\end{lstlisting}
\caption{PendSV and SysTick Interrupt Handler Models}
\label{fig:pendsv-systick}
\end{figure}

Our \promela model follows the FreeRTOS ARM Cortex-M4 port to specify
the scheduler in interrupt handler models for PendSV and SysTick
(Figure~\ref{fig:pendsv-systick}). Both interrupt handler models use
\texttt{IRQ(PID$_\mathit{id}$)} for interrupt
requests and exception entry. When the PendSV interrupt handler
model is executed, a task identification number is chosen by
\texttt{NextTaskId()}. The chosen identification number then
replaces the identification number on the top of stack. In
\texttt{ExpReturn()}, the variable \texttt{EP} is set to the chosen
identification number on stack. The elected task model can continue
its execution.

The SysTick interrupt handler model is similar
(Figure~\ref{fig:pendsv-systick}). When the interrupt handler model is
taken and preemptive scheduling is enabled, it conditionally triggers
the PendSV interrupt with \texttt{SET\_PENDING(PID$_\mathtt{PendSV}$)}.
If it is triggered, the PendSV interrupt presently has not the priority
and will keep pending. The SysTick interrupt handler model then 
executes \texttt{ExpReturn()}. The PendSV interrupt handler model will
be taken directly by tail chaining in our interrupt handler models. 
A task identification number can then be chosen by the PendSV
interrupt handler model as in real hardware.

In our model, the SysTick interrupt is triggered with
\texttt{SET\_PENDING(PID$_\mathtt{SysTick}$)}.
Since the \promela language is timeless, our model cannot trigger
the SysTick interrupt periodically. Rather, our model
non-deterministically triggers the interrupt.
Effectively, the SysTick interrupt is triggered arbitrarily in our
formalization. This is a simple but useful abstraction in our model.
The conservative abstraction ensures that all SysTick interrupt
sequences in real world are subsumed in our model. If there is any
failure among all real interrupt sequences, it will be exposed in our
model. On the other hand, not all interrupt sequences in our model are
real. An error found in the model can be spurious. It has to be
validated by corresponding failures in real hardware.

In cooperative scheduling, a task calls the FreeRTOS yield function to
release the processor. The yield function simply triggers the PendSV
interrupt to elect a task in the PendSV interrupt handler. It is
straightforward to define the FreeRTOS yield function in our model:
\[
\small
\texttt{\#define yield($id$) AWAIT($id$, SET\_PENDING(PID$_\mathtt{PendSV}$))}
\]

\section{Task Synchronization}
\label{sec:delay:suspend}

In addition to task scheduling, the FreeRTOS scheduler also provides
basic functions for task synchronization. More concretely, a task can be
\emph{delayed} for a specified duration; it can also be \emph{suspended}
indefinitely. When a task is delayed, it is moved to a delay queue and
hence cannot be scheduled for execution. Similarly, a suspended task
is moved to a suspended queue and thus ineligible for scheduling. When
its delay duration expires, a delayed task is removed from the delay
queue and can be scheduled for execution. When a suspended task is
resumed by the running task, it is removed from the suspended queue and
ready for scheduling.

In the FreeRTOS ARM Cortex-M4 port, basic task synchronization
functions are implemented by the PendSV and SysTick interrupt
handlers as well. The SysTick interrupt handler checks if any task in
the delay queue has expired its duration periodically. If so,
the interrupt handler removes such tasks from the delay
queue. In contrast, suspended tasks are removed from the suspended queue
when they are resumed by the running task.
If preemptive scheduling is disabled, the
running task continues its execution until it yields the processor.

If preemptive scheduling is enabled, the PendSV
interrupt is triggered when the tasks removed from the delay or suspended
queues have the priority over the running task.
The FreeRTOS scheduler elects a task with the highest
priority for execution. A previously delayed or suspended task will
continue its execution; and the running task will be preempted if it
does not have the priority.

To simplify the boundary condition where all tasks are delayed or
suspended, FreeRTOS creates an \emph{idle} task. The idle task
has the lowest priority and cannot be delayed nor suspended. It can
also be configured to yield the processor or not. If the idle task
should yield, it yields the processor to the next scheduled task immediately.
Otherwise, the idle task loops until it is preempted.

Our model for basic task synchronization follows the FreeRTOS
Cortex-M4 port as well (Figure~\ref{fig:pendsv-systick}). The SysTick
interrupt handler model checks if any delayed task has expired delay
duration. Recall that our formal model is timeless. Delay duration
cannot be formalized exactly. We therefore formalize delay duration by
counters. When a task model is delayed, a counter is set. When the
SysTick interrupt handler model is executed, it decreases counters
of all delayed task models by one. A counter of a delayed task model
is expired if it reaches zero. When their counters are expired, the
SysTick interrupt handler model removes such tasks from the delay queue.
In order to resume suspended tasks, the running task model removes such tasks
from the suspended queue and prepares them for scheduling.
If preemptive scheduling is enabled and the removed tasks have the priority,
the PendSV interrupt is moreover triggered with
\texttt{SET\_PENDING(PID$_\mathtt{PendSV}$)}.

Whether the idle task should yield has impacts on the FreeRTOS
scheduler. The idle task is also formalized in our
model. Our idle task model can also be configured to yield the
processor whenever possible. 

\hide{
Task delaying and suspending are also implemented by the PendSV and
SysTick interrupts. The SysTick interrupt handler checks if the
delay duration of a delayed task has expired. It then activates the
delayed task for scheduling. Similarly, a suspended task is activated
for scheduling when other tasks resume it. If task preemption
is specified in the current scheduling policy and the active tasks have
priority no less than the running task, the PendSV software
interrupt is triggered. The FreeRTOS scheduler will be called to elect
the next task for execution in the PendSV interrupt handler. The
elected task will proceed after the exception return phase.

In order to prevent all tasks from becoming \bred{inactive},
FreeRTOS always creates the \emph{Idle} task. The task has the lowest
priority and is always active. When the preemptive scheduling is specified,
programmers can configure whether the idle task should yield the processor
or not. If it should yield, the idle task minimize the processing time.
Otherwise, the idle task loops until it is preempted.

In our model, a task is delayed when its delay counter is set to
a positive integer. Note that the integer is bounded to the data size
of the delay counter. Since the delayed task
is \bred{inactive}, it will be scheduled after it becomes active.
It is the SysTick interrupt handler model that activates the delayed tasks.
Every time the SysTick interrupt handler is executing, it decreases
all the positive delay counters by one and then activates
the delayed tasks with the zero delay.
}

\section{Thread-Safe Data Structures}
\label{sec:thread-safe-data}

In addition to basic task synchronization, FreeRTOS also provides
thread-safe data structures for message passing and advanced
synchronization among tasks. A thread-safe structure consists of its
data and a waiting task queue.
A task can modify a thread-safe structure immediately when its
data are ready. Otherwise, the task will be \emph{blocked}. When a
task is blocked, it is moved to the waiting task queue of the
thread-safe structure for specified duration. Different
from basic task synchronization, tasks can be unblocked when its
duration is expired or when the data become ready. It is a failure if
the duration of a blocked task is expired before the data are ready.

\hide{
When a task attempts to access the buffer beyond the boundary,
the task needs to wait until the buffer is available. In other words,
the task is blocked on the thread-safe queue. A task is blocked when
it is put into the waiting queue. The blocked task is either delayed
for a programmer-specified duration or suspended indefinitely.
A task is unblocked when it expires the delay duration.
Alternatively, a task is also unblocked when other tasks sharing
the same thread-safe queue makes the buffer available to
the task in the waiting queue.
The unblocked task continues to access the buffer. If the buffer is
still unavailable and the specified duration has not expired, the task
is blocked again. Otherwise, the synchronization outcome is returned.
}

FreeRTOS implements thread-safe queues for message passing. 
A thread-safe queue contains a bounded buffer as its data. The
capacity of the buffer is specified by programmers. The buffer is
ready for modification when it is neither empty nor full. A task will
be blocked when it adds to a full buffer or removes from an empty
buffer. FreeRTOS provides two functions for thread-safe queues.
The \texttt{Send($\mathit{delay}$)} function inserts a
message into the buffer; the \texttt{Receive($\mathit{delay}$)}
function removes a message from the buffer. The parameter
$\mathit{delay}$ specifies the duration.
Consider, for instance, a sender is blocked by sending
a message to a full buffer. When a message is removed from the buffer,
the sender will be unblocked immediately even before its duration
expires. If the 
buffer remains full during the specified duration, the 
sender will be unblocked with a failure. Particularly,
\texttt{Send(0)} returns a failure immediately if the thread-safe
queue is full at the time of invocation. Similarly, \texttt{Receive(0)}
returns a failure if the queue is empty at the time of invocation.

\hide{
 When the buffer is full, \texttt{Send($delay$)} 
acts conditionally. If $delay$ is specified as zero, the function returns
synchronization failure. Otherwise, the sender is blocked.
Similarly,  When the buffer is empty, \texttt{Receive($delay$)} blocks
the receiver conditionally.

}

\hide{
To pass messages between tasks, FreeRTOS provides thread-safe
queues. A thread-safe queue is implemented by a bounded buffer and
shared by a sender and a receiver. The \texttt{Send($delay$)} function inserts
a message into the buffer. When the buffer is full, \texttt{Send($delay$)}
blocks the sender for a programmer-specified duration $delay$ or returns
synchronization failure if $delay$ is specified as zero.
Similarly, the \texttt{Receive($delay$)} function removes a message from
the buffer. When the buffer is empty, \texttt{Receive($delay$)} blocks
the receiver for a programmer-specified duration or returns
synchronization failure if $delay$ is specified as zero.
Note that the sender and receiver tasks are \emph{blocked} on a thread-safe queue.
Consider, for instance, a sender is blocked for a duration by sending
a message to a full buffer. Because a message can be removed from the
buffer at any time, the sender may be unblocked before the specified
duration is expired. Only when no message is removed during the
specified duration, will the sender be unblocked after the duration
has expired.
}

FreeRTOS also offers thread-safe locks. A thread-safe lock uses
a counter as its data. Programmers can also initialize the counter.
Two functions are provided for thread-safe
locks.  The \texttt{Give()} function increments the counter
without blocking. If the counter is not zero, the
\texttt{Take($\mathit{delay}$)} function decrements the counter by
one. Otherwise, the calling task is blocked for the duration specified
by $\mathit{delay}$. Particularly, \texttt{Take(0)} returns a failure
immediately when the thread-safe lock is zero at the time of
invocation.

Thread-safe locks are used to implement mutexes or semaphores for task
synchronization. For mutexes, tasks acquire a lock with
the \texttt{Take($\mathit{delay}$)} function. Not until the lock owner
releases the lock with the \texttt{Give()} function can another task
owns the lock. For semaphores, locks can be taken by a task and
released by either a task or an interrupt handler.

\hide{
The thread-safe queue for data synchronization is applied to emulate
the mutex or the semaphore. When the queue is emulating the mutex,
the taker and the giver should be the same task. Meanwhile, both of
the \texttt{Take($delay$)} and \texttt{Give()} functions
enable the priority
inheritance mechanism. When the queue is emulating the Semaphore,
the taker and the giver is beyond the same task.
}

\hide{
Another useful utility is sharing locks among tasks. In this utility,
a counter replaces the bounded buffer to represent the usage of locks.
Programmers first specify the number of available locks. A task takes
one lock by calling the \texttt{Take($delay$)} function. If no lock is
available, the task is blocked for a programmer-specified duration or
returns failure if $delay$ is specified as zero.
Similarly, a task returns a lock with the \texttt{Give()} function.
The \texttt{Give()} function does not block tasks.
It returns the synchronization outcome immediately.
}

Since thread-safe structures are widely used in FreeRTOS applications,
they are also formalized in our models. Thanks to our ARM Cortex-M4
interrupt model and FreeRTOS scheduler model, our thread-safe
structure models mostly follow the FreeRTOS ARM Cortex-M4 port. 

\section{Applications}
\label{sec:applications}

To illustrate its features and demonstrate recommended programming
styles, FreeRTOS provides example applications.
Particularly, mutexes and semaphores are used
for task synchronization. Thread-safe queues are also found in several
applications for message passing. 

\hide{
In addition to its kernel source codes, FreeRTOS provides a number of
example applications to illustrate its features. These applications
demonstrate FreeRTOS features and are often considered as the
recommended programming style for the real-time operating system.
Indeed, several example applications illustrate task synchronization
mechanisms in FreeRTOS. Since task synchronization mechanisms are most
relevant to the scheduler and its policies, we explain how to build
formal models for these applications in this section.
}

\subsection{Mutexes and Semaphores}
\label{sec:application:mutex}
\textit{Recmutex} is an example application to illustrate
priority inheritance in mutexes. Three tasks with different priorities
are created in the application. They also share a mutex. Initially,
the mutex is taken by the task with the high priority. After it
releases the mutex, the task with the high priority suspends
itself. The mutex is then taken by the task with the medium
priority. Similarly, the task with the medium priority suspends itself
after the mutex is released. The mutex is thus taken by the task with
the low priority. Before releasing the mutex, the running task
resumes the others. Because of priority inheritance, the priority of the
running task should be raised. After the mutex is released, the
running task recovers its low priority.

\hide{
A mutex is shared by them. Obviously, the mutex is first taken by
the highest priority task. After the mutex is released, the highest
priority task suspends itself. The middle priority task performs
the same process as the highest priority task. When the lowest priority task
finally takes the mutex, it subsequently resumes the middle and the highest
priority tasks. Because of the priority inherit mechanism, the lowest
priority task expects its priority will be raised by the other two tasks.
Once the property is checked, the lowest priority task will release the mutex.
In the end, the lowest priority task retrieves its original priority.
}

\textit{Semtest} uses semaphores for task synchronization. The
application contains two pairs of tasks. The tasks in each pair share
a binary semaphore with its counter initialized to one. Both try to
acquire the semaphore by calling \texttt{Take($\mathit{delay}$)}. When
a task acquires the semaphore, it will release the semaphore with the
\texttt{Give()} function. The two task pairs however use different
strategies to acquire semaphores. In the first pair, both tasks call
\texttt{Take(0)} to acquire the semaphore. Since $\mathit{delay}$ is
zero, no task will be blocked. Rather, a failure is returned to the
task without semaphore immediately. The task without semaphore will
yield the processor for the next attempt. In the second task pair,
both tasks call the \texttt{Take($\mathit{delay}$)} with a non-zero
$\mathit{delay}$. The task without semaphore is hence blocked. It will
be unblocked when the semaphore is released.

\hide{
is an example application of semaphores.
It consists of two tasks pairs. Each pair shares a semaphore.
This semaphore has one lock (known as the binary semaphore).
In each pair, two tasks compete with each other for
the semaphore with \texttt{Take($delay$)}.
When one task successfully takes the semaphore, the other should
wait until the semaphore is released with \texttt{Give()}.
In \textbf{Semtest}, the two task pairs have different strategies to
wait for a released semaphore.
In the first pair, the $delay$ parameters of the two \texttt{Take($delay$)}
functions are specified to zero. This means that the two tasks will not
be blocked on the semaphore. If synchronization failure occurs,
the task with no semaphore yields the processor for the next attempt.
In the second pair, the $delay$ parameters of the two \texttt{Take($delay$)}
functions are specified to a duration. When the semaphore is taken by
one of the tasks,
the other will be blocked on that semaphore.
}

\subsection{Queues}\label{sec:BlockQ}


The \textit{BlockQ} application demonstrates how to use thread-safe
queues. The application consists of three task pairs. Each task pair
contains a producer task and a consumer task. The producer task sends
consecutive numbers through a thread-safe queue with the
\texttt{Send($\mathit{delay}$)} function. The consumer task receives
numbers from the queue with \texttt{Receive($\mathit{delay}$)}.
Priorities and $\mathit{delay}$ parameters are different in each
task pair.

In the first pair, the producer task has the high priority but the 
consumer task has the low priority. The producer task also calls
\texttt{Send($\mathit{delay}$)} with a non-zero $\mathit{delay}$ but
the consumer task calls \texttt{Receive(0)} without blocking. In the
second pair, the producer task has the low priority and the consumer
task has the high priority. The producer task calls the non-blocking
\texttt{Send(0)} function. The consumer task on the
other hand calls \texttt{Receive($\mathit{delay}$)} with a non-zero
$\mathit{delay}$. Finally, both producer and consumer tasks have the
low priority and invoke the thread-safe queue functions with non-zero
delay. 

Tasks in \textit{BlockQ} behave very differently in different
scheduling policies. Consider the first task pair where the producer
task has a higher priority than the consumer task. Suppose the
producer task is blocked by \texttt{Send($\mathit{delay}$)} and then
unblocked by the consumer task's \texttt{Receive(0)}. In
preemptive scheduling, the consumer task is preempted by the FreeRTOS
scheduler and the producer task continues its execution. The queue
will not be empty when the consumer task resumes its execution later.
In cooperative scheduling, the consumer task is not preempted when it
calls \texttt{Receive(0)}. If the consumer task never yields, the
producer task will not execute. The queue hence will become empty.
The consumer task will then be blocked by \texttt{Receive(0)}
eventually. No progress can be made in the first task pair. 
To ensure progress, low-priority tasks in \textit{BlockQ} always
yield in cooperative scheduling.

\hide{
In our \promela models, \textbf{BlockQ} is modeled by six execution
units. Each execution unit specifies a task in the application. Using
our formal model for thread-safe queues, it is straightforward to
write formal models for \textbf{BlockQ}. Our \textbf{BlockQ} model
moreover adds \texttt{Yield($id$)} in low-priority execution units
when the cooperative scheduling policy is enabled. It follows
the \textbf{BlockQ} application pretty accurately.
}

\paragraph{}
Using our thread-safe data models, it is almost straightforward to
build models for \textit{Recmutex}, \textit{Semtest}, and
\textit{BlockQ}. We have indeed constructed formal models for several
FreeRTOS example applications such as \textit{PollQ}, \textit{QPeek},
\textit{Dynamic}, \textit{Countsem}, and \textit{GenQTest}. These
applications are selected because they illustrate task synchronization
and thread-safe structures in FreeRTOS. They are useful in our
formal analysis of the FreeRTOS scheduler on ARM Cortex-M4 processors.

\section{Formal Properties}
\label{sec:formal-properties}

It is impossible to analyze the FreeRTOS scheduler formally without
formal properties. Such properties nevertheless are not always
obvious. In real systems like FreeRTOS, high-level properties cannot
be established without caveats. Without necessary provisions, formal
properties could be contrived or even meaningless. In order to avoid
contrived formal properties, we verify assertions in
FreeRTOS example applications.

FreeRTOS developers annotate example applications with many
assertions for testing. If a particular task schedule results in an
assertion failure, it indicates an unintended behavior in an example
application. In our analysis, we aim to prove the absence of assertion
errors among all task schedules in our formal models. Since assertions
are annotated by FreeRTOS developers, they are not contrived.
Our models moreover simulate the ARM Cortex-M4 interrupt mechanism,
the FreeRTOS scheduler, task synchronization, and thread-safe
structures. Assertion errors found in our models likely
correspond to assertion failures in FreeRTOS example applications. Our
formal analysis is realistic as well.

Not all assertions are similar however. To organize our presentation,
we classify assertions in FreeRTOS example applications into two
categories. Intuitively, an assertion specifies a safety property if
it indicates that a bad event should never happen; an assertion
specifies a liveness property if it indicates that a good event should
always happen.

\hide{
Our models moreover simulate the ARM Cortex-M4 interrupt mechanism,
the FreeRTOS scheduler, task synchronization, and thread-safe
structures, task schedules are pretty realistic in our
models. Accuracy of our formal analysis is therefore greatly
improved. 
Assertion errors
in our analysis likely correspond to assertion failures in FreeRTOS
example applications. We will examine assertion errors and reproduce
them in real hardware. FreeRTOS developers can then evaluate assertion
failures 
}

\hide{
In their example applications, FreeRTOS developers specify expected
program behaviors with assertions. In the C standard library, the
\texttt{assert($\mathit{bexp}$)} function reports an assertion failure
if the Boolean expression $\mathit{bexp}$ is false at runtime. Such
assertions are often used for testing. Any assertion failure is seen
as an error in testing. Intuitively, these assertions are properties
about applications. They are easily specified in our formal models
because \promela also supports assertions. We briefly describe them in
this section.
}

\subsection{Safety}

It is straightforward to specify safety properties with
assertions. Programmers only need to write a Boolean expression deemed
to be true in an assertion. In FreeRTOS example applications, the
following safety properties are found:
\begin{enumerate}
\item[(S0)] If a task is delayed for synchronization with other tasks,
  other tasks must finish before the delay duration expires.
  \label{S0}
\item[(S1)] If a task is blocked by a thread-safe data,
  the data must be ready when the task is unblocked.
  \label{S1}
\item[(S2)] If a task expects a thread-safe data to be ready,
  the data must be ready.
  \label{S2}
\item[(S3)] Messages received through a thread-safe queue must preserve
  their order.
  \label{S3}
\item[(S4)] Mutexes and binary semaphores must ensure mutual exclusive
  execution of critical sections.
  \label{S4}
\item[(S5)] Frequencies of \texttt{Take($delay$)} and \texttt{Give()}
  must be equal.
\hide{ the same when more than one lock are shared by mutexes or
  semaphores. }
  \label{S5}
\item[(S6)] A low-priority task must inherit priorities when its mutex
  was taken by tasks with higher priorities.
  \label{S6}
\end{enumerate}

Property (S0) checks if basic task synchronization is used properly.
When a task is delayed, the delayed duration must be sufficient for
other tasks to finish their works. Property (S1) checks if thread-safe
data are implemented correctly. When a task is unblocked before its
duration expires, the thread-safe data must be ready.
Property (S2) is a special case of property (S1) where the parameter
$delay$ is zero. Property (S3) checks that messages are delivered in
order by thread-safe queues. Properties (S4) and (S5) check mutexes
and semaphores are implemented correctly. Finally,
property (S6) checks whether priority inheritance is implemented.

Not all properties are required in every
application. Table~\ref{tab:specs} shows the safety properties 
specified in the eight FreeRTOS example applications.

\begin{table}
  \small
  \caption{Properties in FreeRTOS Applications}
  \begin{tabular}{|l|c|c|c|c|c|c|c|c|}
    \hline
    \multirow{2}{*}{} & \multicolumn{7}{c|}{Safety} & \multirow{2}{*}{Liveness} \\
    \cline{2-8}
                & S0 & S1 & S2& S3 & S4 & S5 & S6 &   \\
    \hline
    \textit{PollQ}       &\cmark&   &\cmark&\cmark&   &   &   &\cmark\\
    \textit{Semtest}     &\cmark&\cmark&\cmark&   &\cmark&   &   &\cmark\\
    \textit{BlockQ}*     &   &\cmark&\cmark&\cmark&   &   &   &\cmark\\
    \textit{QPeek}*      &   &\cmark&\cmark&\cmark&   &   &   &\cmark\\
    \textit{Dynamic}*    &\cmark&   &\cmark&\cmark&\cmark&   &   &\cmark\\
    \textit{Countsem}*   &   &   &\cmark&   &   &\cmark&   &\cmark\\
    \textit{Recmutex}*   &   &\cmark&   &   &\cmark&\cmark&\cmark&\cmark\\
    \textit{GenQTest}*   &   &\cmark&\cmark&\cmark&\cmark&   &\cmark&\cmark\\
    \hline
  \end{tabular}
  {\footnotesize * The application has additional \texttt{yield($id$)}
    commands when cooperative scheduling is specified. }
  \label{tab:specs}
\end{table}

\subsection{Liveness}\label{sec:liveness_property}

If a task does nothing, no bad event can happen. The task thus
satisfies all safety properties. To avoid such vacuous
safety, liveness properties are needed. Indeed, FreeRTOS developers
write assertions to ensure tasks are making progress. To
check progress by assertions, a task maintains a counter. The task
increments the counter when its job is finished. Another task is added
to 
check the counter periodically. It is an assertion failure if the
counter remains unchanged between checks.

Adding a task to check progress is fine, but the added check task does
not really contribute to the work of application. Precious energy and 
processor cycles are consumed by the check task. More importantly, the
check task needs to be scheduled. Task schedules would be different should
the check task be removed from its application. If an application is
tested with a check task, it needs to be shipped with the check
task in the final product. Otherwise, test results are debatable
because task schedules necessarily change without the check task.

\hide{
Informally, liveness properties specify if certain good events must
happen. Task progression can be seen as a liveness property. 
One can check if tasks are always making progress with
assertions. With our formal models, more sophisticated
liveness properties can be checked. More specifically, we would like
to know if any task can stop making progress from a certain point on.

The difference between always making progress and no progress at all
is significant. Ideally, all user tasks should make progress
always. It however may still be tolerable if a user task fails to
perform once in a while. As long as all user tasks can make some
progress eventually, it might be acceptable but not preferable in some
circumstances. On the other hand, it is never acceptable if a user task
stops making progress. Such phenomena often indicate errors in
applications or even schedulers in operating systems.
}

Such a dilemma is resolved in our formal analysis easily. Instead of
checking progress in a task, we specify the liveness property with an
LTL formula. Since the logic formula is not an
execution unit, it has 
no impact on task schedules. Actually, the formula is not even a part
of our formal models. Model behaviors cannot be changed. Formal models
allow us to check progress without adding check tasks. Our analysis is
valid for final products without check tasks.

\hide{
To check if any user task may stop making progress, we first use the
LTL formula
\[
    \Box\Diamond\mathit{Loc}_{1}\wedge
    \Box\Diamond\mathit{Loc}_{2}\wedge\cdots\wedge
    \Box\Diamond\mathit{Loc}_{n}
\]
where $\mathit{Loc}_{i}$'s are locations where user tasks perform
their jobs. The formula specifies that each location $\mathit{Loc}_i$
is visited infinitely many times. If the formula holds in our models,
then all user tasks perform their jobs infinitely many times for all
possible traces. The model checker \spin nevertheless finds a deviant
trace. After examining the trace, we notice that the SysTick interrupt
stops running in the trace. Since the SysTick interrupt stops,
FreeRTOS scheduler cannot schedule user tasks. User tasks of course do
not perform their jobs regularly. Such a trace however is
spurious. The SysTick interrupt must be triggered infinitely many
times in reality.
}

Precisely, let $\mathit{Loc}_{\mathit{SysTick}}$ be the location
triggering the SysTick interrupt and $\mathit{Loc}_i$ the location
where task model $i$ finishes its job for $1 \leq i \leq n$. Consider
the LTL formula:
\[
  \Box\Diamond\mathit{Loc}_{\mathit{SysTick}}\rightarrow(
    \Box\Diamond\mathit{Loc}_{1}\wedge
    \Box\Diamond\mathit{Loc}_{2}\wedge\cdots\wedge
    \Box\Diamond\mathit{Loc}_{n}.
  )
\]
Informally, the formula states that all tasks finish their jobs
infinitely many times if the SysTick interrupt is triggered infinitely
many times. In our formal models, SysTick interrupts represent the
progression of time. If the LTL formula is
satisfied in our models, it means that all task models must finish
their jobs infinitely often as time progresses. In other words, no
task can stop making progress indefinitely. We verify this liveness
property in place of assertions from check tasks in our formal
analysis. The liveness property is required for all FreeRTOS
application models (Table~\ref{tab:specs}).

\section{Verification Results}
\label{sec:evaluation}

\hide{
With our formal models for the ARM Cortex-M4 interrupt mechanism, 
scheduler, task synchronization, and FreeRTOS example applications,
}
For each scheduling policy, we use the model checker \spin to verify
properties shown in Table~\ref{tab:specs}. The model checker first
verifies whether there is any assertion error for
safety properties in an application model. After checking safety
properties, \spin is used again to verify the liveness property on the
application model. In our experiments, we use \spin 6.5.1 on an Ubuntu 
20.04 server with two 3.2GHz octa-core CPUs and 512GB RAM.

Table~\ref{tab:results} gives the verification results for
safety properties in eight example applications under three scheduling
policies. If all safety properties in an application are satisfied, the
verification time (in seconds) is shown. If not, the failed property
is shown with a cross mark in the table. For the liveness property,
verification time is shown if an application satisfies the
property. A cross mark represents that an application does not satisfy the
liveness property.


\begin{table*}
  \centering\small
  \caption{Verification Time in Seconds}
  \begin{subtable}[t]{.3\textwidth}
    \centering
    \caption{Cooperative Scheduling}
    \label{tab:results:cooperative}
    \begin{tabular}{|l|r|r|}
      \hline
      & Safety & Liveness \\
      \hline
      \textit{PollQ}       & 2.9   & 59.3\\
      \textit{Semtest}     & 0.2   & \xmark\\
      \textit{QPeek}*      & 0.1   & 2.4\\
      \textit{Recmutex}*   & 12.5  & 428.0\\
      \textit{Countsem}*   & < 0.1 & 0.6\\
      \textit{GenQTest}*   & 1.7   & 75.9\\
      \textit{Dynamic}*    & 16.9  & 340.0\\
      \textit{BlockQ}*     & 3.1   & 424.0\\
      \hline
    \end{tabular}
  \end{subtable}
  \begin{subtable}[t]{.3\textwidth}
    \centering
    \caption{Preemptive w/o Time Slicing}
    \label{tab:results:without-time-slicing}
    \begin{tabular}{|l|r|r|}
      \hline
      & Safety & Liveness\\
      \hline
      \textit{PollQ}    & 5.5   & 145.0\\
      \textit{Semtest}  & 142.0 & \xmark\\
      \textit{QPeek}    & < 0.1 & 1.9\\
      \textit{Recmutex} & 9.8   & 375.0\\
      \textit{Countsem} & < 0.1 & \xmark\\
      \textit{GenQTest} & 0.1   & \xmark\\
      \textit{Dynamic}  & 0.4   & \xmark\\
      \textit{BlockQ}   & 5.2   & 1010.0\\
      \hline
    \end{tabular}
  \end{subtable}
  \begin{subtable}[t]{.3\textwidth}
    \centering
    \caption{Preemptive w/ Time Slicing}
    \label{tab:results:with-time-slicing}
    \begin{tabular}{|l|r|r|}
      \hline
      & Safety & Liveness\\
      \hline
      \textit{PollQ}    & 7.7  & 210.0      \\ 
      \textit{Semtest}  & 934.0& \xmark \\ 
      \textit{QPeek}    & 0.1  & \xmark \\
      \textit{Recmutex} & 57.7 & \xmark \\ 
      \textit{Countsem} & 3.7  & \xmark \\
      \textit{GenQTest} & 332.0& \xmark \\
      \textit{Dynamic}  & S0 \xmark & \xmark \\ 
      \textit{BlockQ}   & S1 \xmark & \xmark \\ 
      \hline
    \end{tabular}
  \end{subtable}

  * The application has additional \texttt{yield($id$)} commands when
  cooperative scheduling is specified.
\hide{
  \begin{tabular}{|l|r|r|c|l|r|r|c|l|r|r|}
    \cline{1-3}\cline{5-7}\cline{9-11}
    \multicolumn{3}{|c|}{\textbf{Cooperative Scheduling}} &&
    \multicolumn{3}{|c|}{\textbf{Preemptive Scheduling without Time Slicing}} &&
    \multicolumn{3}{|c|}{\textbf{Preemptive Scheduling with Time Slicing}} \\
    \cline{1-3}\cline{5-7}\cline{9-11}
    & Safety & Liveness &&\hspace{8em} & Safety & Liveness &&\hspace{8em} & Safety & Liveness \\
    \cline{1-3}\cline{5-7}\cline{9-11}
    PollQ       & 2.9   & 59.3      && PollQ    & 5.5   & 145.0      && PollQ    & 7.7  & 210.0      \\ 
    Semtest     & 0.2   & \xmark && Semtest  & 142.0 & \xmark && Semtest  & 934.0& \xmark \\ 
    QPeek*      & 0.1   & 2.4       && QPeek    & < 0.1 & 1.9        && QPeek    & 0.1  & \xmark \\
    Recmutex*   & 12.5  & 428.0     && Recmutex & 9.8   & 375.0      && Recmutex & 57.7 & \xmark \\ 
    Countsem*   & < 0.1 & 0.6       && Countsem & < 0.1 & \xmark && Countsem & 3.7  & \xmark \\
    GenQTest*   & 1.7   & 75.9      && GenQTest & 0.1   & \xmark && GenQTest & 332.0& \xmark \\
    Dynamic*    & 16.9  & 340.0     && Dynamic  & 0.4   & \xmark && Dynamic  & S0 \xmark & \xmark \\ 
    BlockQ*     & 3.1   & 424.0     && BlockQ   & 5.2   & 1010.0     && BlockQ   & S1 \xmark & \xmark \\ 
    \cline{1-3}\cline{5-7}\cline{9-11}
    \multicolumn{11}{l}{\footnotesize
      \begin{tabular}[x]{@{}l@{}}
        * The application is different from the others with the preemptive scheduling policy.
      \end{tabular}
    }
  \end{tabular}
}
  \label{tab:results}
\end{table*}

\subsection{Analysis of Safety Properties}\label{sec:safety}
\hide{
For applications satisfying all safety
properties, \spin finishes in 16 minutes and uses up to 56GB
memory. For disproved properties, \spin reports error
traces in 1 minutes with 10GB memory.
}

Almost all applications satisfy their safety properties. \spin
finishes the verification with at most 56GB of memory in 16
minutes. For failed safety properties, the model checker also reports
error traces with 10GB memory in 1 minute.

\hide{ 
Consider the demonstration of priority
inheritance in \S\ref{sec:IntSemTest}. Because the preemptive scheduling
is closed, the lower priority task will not be preempted after it resumes
the higher priority task. This makes the lower priority task continues
without inheriting the high priority and eventually violates (S6).

This disproved result only says that \textbf{IntSemTest} cannot be executed
with the cooperative scheduling. We easily reproduce this disproved result
by compiling the application with the cooperative scheduling policy. Once
the binary file is flashed to the Discovery kit, the application immediately
informs the violation of (S6) through onboard LEDs.
}

Under preemptive
scheduling with time slicing, \spin finds that the application models
\textit{Dynamic} and \textit{BlockQ} violate safety properties
(S0) and (S1) respectively. In error traces reported 
by \spin, we find that a task may not execute even though
it is scheduled by the FreeRTOS scheduler. To see how it happens,
consider the SysTick interrupt triggers while the PendSV interrupt
handler model is running. Since both interrupts have the same 
priority, the SysTick interrupt is pending until the PendSV interrupt
handler model finishes. Recall that the PendSV interrupt handler model
calls the scheduler model to elect a task model for execution.
Let us call the elected task model as the \emph{victim}.
The victim task model is scheduled to execute after the exception return
macro. However, the SysTick interrupt is still pending. Due to tail
chaining, the SysTick interrupt handler model will execute before the
victim task. In the time slicing policy, the SysTick interrupt handler
mode will trigger the PendSV interrupt to schedule a task model. The
scheduler model incorrectly believes the victim task model has 
used up its time slice and chooses another task model for execution.
The victim task model hence misses its time slice for execution.
In error traces, the victim task model repeatedly misses its time slice
and hence cannot prepare the thread-safe queue shared with
another blocked task. When the blocked task expires its duration,
the shared thread-safe queue is still not ready. Eventually,
the models \textit{Dynamic} and \textit{BlockQ} violate
safety properties (S0) and (S1) respectively.

\hide{
Not all properties are proved in our formal verification. Some
applications, for instance, only work correctly for specific
scheduling policies. Corresponding formal models may not satisfy all
properties under all scheduling policies during verification. Our
formal models on the other hand are not FreeRTOS applications. False
alarms may arise due to inherent inaccuracy in formal models.
Consequently, a property disproved by verification is not necessarily a
real problem. In order to justify our verification results, we
reproduce disproved properties on the STM32F429I-DISC1 board
with an ARM Cortex-M4 processor. More concretely, an on-board LED
is used to indicate assertion failures in running FreeRTOS example
applications. We instrument FreeROTS applications to observe assertion
failures for disproved properties. 
}

Although assertion errors are found in our formal analysis, they are
not necessarily failures in reality. It is important to recall that
our application models are not FreeRTOS example applications. During
model construction, abstraction and simplification are indispensable
for effective formal analysis. For instance, the SysTick interrupt is
not triggered periodically in our formal models because the \promela
modeling language is timeless. Error traces found by \spin are only
realistic but never real. It is just as important to reproduce
assertion failures in real hardware for assertion errors found in
formal analysis. To this end, we install the FreeRTOS ARM Cortex-M4
port on the STM32F429I-DISC1 board with an ARM Cortex-M4 processor and
modify FreeRTOS example applications to reproduce \spin error traces
on the board. If
an assertion failure does occur, an on-board LED will flash with high
frequency to indicate the failure.

The failed safety properties in \textit{Dynamic} and
\textit{BlockQ} under preemptive scheduling with time slicing are
successfully reproduced on STM32F429I-DISC1.
Consider the safety property (S1) in \textit{BlockQ}.
We add a task to the example application. When the
added task is scheduled for execution, it runs for the time slightly
shorter than the SysTick period and then yields.
After the added task yields, the FreeRTOS scheduler will
choose, for example, a producer task in \textit{BlockQ}. 
The elected producer task will be the victim. While the FreeRTOS
scheduler is electing the victim task, the SysTick interrupt is
triggered but remain pending. As described above, the victim task will
be preempted before it executes. An assertion failure in the victim
producer task is observed.

In reality, the SysTick interrupt may not be triggered shortly after
the added task yields. The added task simply repeats itself whenever
it is scheduled for execution. The assertion failure will be observed
eventually. The failed safety property (S0) in \textit{Dynamic} is
reproduced similarly. Two assertion failures are found by our formal
analysis successfully.

After reproducing the failures, we find that similar pattern had been
independently exploited. Tsafrir et al.~\cite{268460} made non-privileged
applications arbitrarily monopolize processors by controlling
processor cycles between two clock ticks. They concluded that
any periodically ticking system at that time is vulnerable to their exploit.
Their exploit and our reproduction are similar in controlling processor
cycles between ticks, but different in the cause of the problem.

\subsection{Analysis of Liveness Property}\label{sec:liveness}

Table~\ref{tab:results} also reports verification results for the
liveness property in all example application models under different
scheduling policies. \spin uses up to 21GB of memory within 17 minutes
for each verification run. Many example
application models do not satisfy the liveness property.


\subsubsection{Liveness under cooperative scheduling}

Only one application model fails to satisfy the liveness property in
Table~\ref{tab:results:cooperative}. The error trace reported by \spin
shows that two of the task models in \textit{Semtest} never yield. Recall
that the first task pair in \textit{Semtest} have zero delay
(Section~\ref{sec:application:mutex}). Since preemption is disabled, 
other task models cannot be scheduled for execution. Both task models
in the first task pair are moved to the waiting task queue. No
progress can be made as time progresses. The liveness property
fails. 

It is easy to reproduce the error on the STM32F429I-DISC1 board.
We configure FreeRTOS to use the cooperative scheduling
policy. The on-board LED indicates an assertion failure without
modifying the FreeRTOS example application.

\subsubsection{Liveness under preemption without time slicing}
\label{sec:liveness:without-time-slicing}

The application models \textit{Semtest}, \textit{Countsem},
\textit{GenQTest}, and \textit{Dynamic} violate the liveness property
under the preemptive scheduling without time slicing
(Table~\ref{tab:results:without-time-slicing}).
After examining their error traces, we find a task model never yields
and other task models are not delayed in each model.
Since time slicing is not enabled, the SysTick
interrupt handler model does not trigger the PendSV interrupt. No task
model will be scheduled for execution. When thread-safe structures
become not ready, never-yielding task models will be moved to waiting
task queues. No progress can be made afterwards. The liveness
property subsequently fails.

It is easy to reproduce assertion failures in
\textit{Semtest}, \textit{Countsem}, and \textit{GenQTest}. After
configuring FreeRTOS with the scheduling policy, assertion failures
in check tasks are observed without any modification.

Most interestingly, \textit{Dynamic} does not have assertion failures
under preemptive scheduling without time slicing. Contrary to our
formal models, recall that additional tasks are used to check progress
in these example applications
(Section~\ref{sec:liveness_property}). These check 
tasks have the highest priority with non-zero delays. The never-yielding
task in \textit{Dynamic} is preempted by its check task periodically;
other tasks will then be scheduled for execution. Progress can still be
made because of the check task in \textit{Dynamic}. To reproduce the
assertion failure in \textit{Dynamic}, we have to change the execution
order of consumer and producer tasks in the example application. After
this simple modification, an assertion failure in the check task is
observed in \textit{Dynamic}.


\subsubsection{Liveness under preemption with time slicing}
\label{sec:liveness:with-time-slicing}

Surprisingly, the liveness property fails in almost all application
models under preemptive scheduling with time slicing. After examining
error traces, the problem in Section~\ref{sec:safety} is observed
again. When the SysTick interrupt is triggered while the PendSV
interrupt hander model is running, recall that a victim task model
will miss its chance of execution. In the extreme scenario, a 
task model can be the victim whenever it is scheduled. The victim task
model will never execute and starve. The liveness property hence
fails.

It can be very tricky to reproduce the starvation on real
hardware.
As a proof of concept, we choose the example application
\textit{Countsem} with two never-yielding tasks to reproduce the
error. The idle task is configured to yield in the
application. Similar to Section~\ref{sec:safety}, we add a new task to
\textit{Countsem}. The added task occupies the ARM Cortex-M4 processor
for a fixed time. It ensures the SysTick interrupt is triggered shortly
after the idle task yields. When the idle task yields, a task is
elected and becomes the victim. The second task will be elected. After
the second task finishes its execution, the added task repeats and
forces the first task to be the victim again.

Incidentally, \textit{PollQ} satisfies the liveness property.
We observe two factors that make \textit{PollQ} immune from the problem
in Section~\ref{sec:safety}.
First, other tasks have higher priorities
than the idle task. This prevents the idle task from preempting the
next task when the idle task should yield the processor.
Second, tasks in \textit{PollQ} delay themselves after synchronization.
Recall that a running task may unblock others through the thread-safe
structure. If the unblocked tasks have the priority,
the running task is preempted. When the preempted
task continues its execution, it then delays itself. The delay can
prevent the executing task from repeatedly preempting the next task.
Based on these factors, we propose a fix for the problem.
The first step is to disable the idle task from yielding the processor.
The second step is to delay the tasks that will definitely yield
the processor by being preempted. The proposal
is not the most efficient fix due to additional delays,
but it can alleviate the problem without modifying the FreeRTOS
kernel code.

\subsection{Discussion}


In our formal analysis, we find three types of assertion failures in
FreeRTOS example applications. Section~\ref{sec:safety} reports
assertion failures where a task may be preempted before its scheduled
execution. Under preemptive scheduling with time slicing, the FreeRTOS
scheduler can be invoked twice by consecutive executions of the PendSV
and SysTick interrupt handlers. The task elected by the first
invocation is preempted by the second invocation before its
scheduled execution. A task therefore may not execute after it is
scheduled. To observe such failures, the PendSV and SysTick
interrupts need to be synchronized. Testing is unlikely to reveal such
failures. We are not aware of any report about such assertion
failures.

The second type of assertion failures are reported in
Section~\ref{sec:liveness:without-time-slicing}. Under preemptive
scheduling without time slicing, never-yielding tasks can lead to
starvation when they use thread-safe structures. In this case, other
tasks cannot be scheduled because never-yielding tasks are
running. When thread-safe structures become not ready, never-yielding
tasks are moved to waiting task queues and applications cannot
progress. The second type of assertion failures can be elusive.
Since FreeRTOS example applications add tasks to check
progress. Because these check tasks change FreeRTOS scheduling, 
starvation may not happen. Even though check tasks do not contribute
to computation actively, applications must be shipped with check tasks
to prevent starvation. This is perhaps the most interesting lesson
learned from our formal analysis.

Section~\ref{sec:liveness:with-time-slicing} reports the third type
of assertion failures. These failures are closely related to the first
type. In the extreme scenario, a task is always preempted before its
scheduled execution. The victim task cannot progress. It is almost
impossible for testing to find such assertion failures. Yet we have
successfully produced it in a FreeRTOS example application with the
help of  our formal analysis.

\section{Related Work}
\label{sec:relate}


Many works had applied formal methods on FreeRTOS.
Chong et al.~\cite{chong2021, web:chong2020} from the Amazon Web Services
team formally verified the FreeRTOS's
task and queue implementation against memory safety such as buffer overflow,
use after free, or NULL pointer dereferences. They propose a process to
check the software in development~\cite{10.1145/3377813.3381347} with
Bounded Model Checking~\cite{10.1007/978-3-540-24730-2_15}.
Chandrasekaran et al.~\cite{6961844} modeled a custom
implementation of multi-core FreeRTOS in \promela. They used \spin to verify
their model against data-race and deadlock. Besides model checking,
some works choose theorem proving. The scheduler~\cite{6269627} and
the list data structure~\cite{7384236} of FreeRTOS were proved
memory correctness with separation logic. Other works modeled
the FreeRTOS scheduler in the B method~\cite{10.1007/978-3-642-10452-7_8}
or the Z notation~\cite{10.1007/s00165-014-0308-9}. Both of the works proved
FreeRTOS task invariants such as the running task always has the top priority.
Divakaran et al.~\cite{10.1007/978-3-319-25423-4_11} verified the FreeRTOS
scheduler with refinement-based method. Specifically, the method not only
checks functional correctness but also proves that
their abstraction in the Z notation is refined by the FreeRTOS
implementation. In automata-based modeling method, Asadollah et al.~\cite{8452035}
applied runtime verification on FreeRTOS. They parsed FreeRTOS' runtime events
to discover concurrent bugs such as deadlock and starvation.
None of the above works considers architecture effects such as tail chaining.
Architecture effects are highly relied on interrupt handling that often
changes the processor context at runtime.

Some works considered hardware interrupts and preemption.
Feng et al.~\cite{10.1145/1375581.1375603} provided a Hoare-logic-like
framework for certifying low-level system programs with interrupts enabled.
Xu et al.~\cite{10.1007/978-3-319-41540-6_4} also provided a framework
for specifying kernel behaviors with preemption and nested interrupts enabled.
They further verified the kernel $\mu$C/OS-II against priority inversion.
Our work has a similar goal, but further involves specific
architecture effects.

As for other verified real-time kernels, de Oliveira
et al.~\cite{deOliveira2019ISORC} modeled thread behaviors in
the Linux PREEMPT\_RT kernel based on automata. In the expansion
of~\cite{deOliveira2019ISORC}, de Oliveira et al.~\cite{deOliveira2019SEFM}
developed a Linux kernel module to keep checking Linux runtime events are
allowed by the automata model. Their method is efficient at
finding mismatching behaviors between the logged events and the model.
Hladik et al.~\cite{hladik:hal-03017661} proposed an integrated tool to
design, verify, and execute a real-time system. Their tool can guarantee that
real-time tasks are executed in time constraint. Real-time tasks are C
functions. They are managed by the tool and executed in a Linux PREEMPT\_RT
kernel at a specific tick frequency.
Andronick et al.~\cite{10.1007/978-3-319-43144-4_4} proved the eChronos
real-time operating system against the scheduling policy that the running task
has the top priority. Other works analyzed response time with
model checking or theorem proving. Guo et al.~\cite{8305936} verified
communications in an in-vehicle network against timed properties.
They abstracted nodes from the network and verified the abstraction with
the timed model checker \textsc{UPPAAL}~\cite{10.1007/BFb0020949}.
Cerqueira et al.~\cite{7557887} provides \textsc{Prosa} for proving
task scheduling and response time. Their framework can be checked by the
\textsc{Coq} theorem prover. Guo et al.~\cite{10.1007/978-3-030-25543-5_28}
applied \textsc{Prosa} to RT-CertiKOS for further analysis.

Besides real-time kernels, notable verified microkernel
seL4~\cite{10.1145/1629575.1629596} is proved functional correctness and
their abstract model is also validated~\cite{10.1145/2491956.2462183}.
CertiKOS~\cite{199344} is a concurrent operating system and supports multicore.
Gu et al. proved that the CertiKOS implementation refines its specifications.
Nelson et al.~\cite{10.1145/3132747.3132748} developed an operating system
kernel that is amenable to automated reasoning using satisfiability modulo
theories solvers. In contrast to the above, we choose to validate
our abstract model by reproducing the reported errors in real hardware.

\section{Conclusion}
\label{sec:conclusion}

We have presented a formal model for the FreeRTOS scheduler and
a number of standard FreeRTOS applications on ARM Cortex-M4 cores.
The standard FreeRTOS applications contains assertions to specify
expected behaviors. By model checking, we find several assertion errors
when we verify some application models under certain scheduling policies.
Those assertion errors are analyzed and reproduced on a physical
development board with the ARM Cortex-M4 core.

Based on our formalization of the ARM Cortex-M4 interrupt mechanism,
the FreeRTOS scheduler model is specified almost naturally. Interrupt
mechanisms in different architectures are also exploited in various
FreeRTOS ports. How to generalize our methodology of formalization
would be an interesting future work for formal analysis of other
FreeRTOS ports.
Specifically, we plan to model FreeRTOS SiFive HiFive1-RevB port to
analyze how RISC-V architecture affects the FreeRTOS scheduler.
Another future work is to model the official distribution of
FreeRTOS symmetric multiprocessing~\cite{web:FreeRTOSSMP} and
consider the effect of memory model.


\section*{Availability}
Our model is available at
\url{https://github.com/kaizsv/FreeRTOS-Spin} and the reproduction is available
at \url{https://github.com/kaizsv/FreeRTOS-Spin-Reproduction}.

\bibliographystyle{plain}
\bibliography{main}

\end{document}